\documentclass[prb,twocolumn,superscriptaddress]{revtex4-2}
\usepackage{amsfonts}
\usepackage{amssymb}
\usepackage{amsmath}
\usepackage{graphicx}
\usepackage{dcolumn}
\usepackage{bm}
\usepackage{units}
\usepackage{soul}
\usepackage{multirow}
\usepackage{CJKutf8}
\usepackage{subfigure}
\usepackage{color}%
\usepackage{url}
\usepackage[colorlinks,linkcolor=blue,anchorcolor=blue,citecolor=blue,urlcolor=blue]{hyperref}
\usepackage{array}
\usepackage{ulem}
\usepackage{appendix}

\begin{document}
\title{Particle-continuum-medium duality of skyrmions}
\author{X. R. Wang}
\email[Electronic address: ]{phxwan@ust.hk}
\affiliation{Physics Department, The Hong Kong University of Science and
	Technology, Clear Water Bay, Kowloon, Hong Kong}
\affiliation{HKUST Shenzhen Research Institute, Shenzhen 518057, China}
\date{\today}
\affiliation{William Mong Institute of Nano Science and Technology, The Hong Kong 
	University of Science and Technology, Clear Water Bay, Kowloon, Hong Kong, China}

\author{X. C. Hu}
\affiliation{Physics Department, The Hong Kong University of Science and
	Technology, Clear Water Bay, Kowloon, Hong Kong}
\affiliation{HKUST Shenzhen Research Institute, Shenzhen 518057, China}
\begin{abstract}
Topological solitons are crucial to many branches of physics, such as 
models of fundamental particles in quantum field theory, information 
carriers in nonlinear optics, and elementary entities in quantum and 
classical computations. Chiral magnetic materials are a fertile ground for 
studying solitons. In the past a few years, a huge number of all kinds 
of topologically protected localized magnetic solitons have been found. 
The number is so large, and a proper organization and classification is 
necessary for their future developments. Here we show that many topological 
magnetic solitons can be understood from the duality of particle and elastic 
continuum-medium nature of skyrmions. In contrast to the common belief that 
a skyrmion is an elementary particle that is indivisible, skyrmions behave 
like both particle and continuum media that can be tore apart to bury 
other objects, reminiscing particle-wave duality in quantum mechanics. 
Skyrmions, like indivisible particles, can be building blocks for cascade 
skyrmion bags and target skyrmions. They can also act as bags and glues to 
hold one or more skyrmions together. The principles and rules for stable 
composite skyrmions are explained and presented, revealing their rich and 
interesting physics.   
\end{abstract}

\maketitle
\section{Introduction} 
Magnetic skyrmions have attracted much attention in recent years for their academic 
interest and potential applications in information technology 
\cite{Bogdanov2001,Rossler2006,Muhlbauer2009,Yu2010,Yu2011,Fert,Nagaosa,Romming}. 
Various aspects of magnetic skyrmions have been extensively and intensively studied 
\cite{Xiansi,Zhou2014,Li,Yuan2016,Jiang,Tian,Onose,Park,Tian,Heinze,news,Woo2016,
Iwasaki2013,Xu,MnSi_anis,size2015,JMMM2,Lenov2016,Braun,size2016,PdFeIr,BandK,
MnSi,gongxin,Kai2017,Yuan2018,Reich2015,Reich2016,Thiaville2013,
Sampaio2013,Lin2013,koshibas2018,Judge2018,Kim2018,Hoshino2018,Woo2018}, 
including characterization, generation, and manipulations. 
Recently, it was realized that only skyrmions with positive formation energy 
are intrinsically circular \cite{paper1,paper2,paper3,paper4}. The natural 
morphology of skyrmions of negative formation energy are stripes of well-defined width. 
In this case, condensed stripy skyrmions, ranging from irregular maze to periodically 
arranged helical states, is the preferred thermodynamic equilibrium state \cite{paper4}. 
Randomly or orderly arranged condensed stripe skyrmions transform into skyrmion crystals 
(SkXs) smoothly and continuously as stripe width increases via material parameter 
engineering \cite{paper5} or as the skyrmion density increases through thermal 
agitations and/or the assistances of external forces \cite{paper3,paper4}. 

With all these advances in skyrmion physics, there are still many unsolved problems. 
For example, other than different elementary skyrmions mentioned above, in the past 
eight years, many localized magnetic structures, topologically different from the 
elementary skyrmions and with arbitrary integer skyrmion numbers, were found \cite
{Leonov14,skyrmionium1,skyrmionium2,skyrmionium3,skyrmionium4,skyrmionium5,skyrmionium6,
skyrmionium7,skyrmionium8,skyrmionium9,skyrmionbag1,skyrmionbag2,skyrmionbag3,
skyrmionbag4,skyrmionbag5,skyrmionbag6,skyrmionbag7,skyrmionbag8,skyrmionbag9,skyrmionbag10,
skyrmionbag11,skyrmionbag12,skyrmionbag13}. 
Skyrmionium, a spin texture of one skyrmion 
inside another larger skyrmion with zero net skyrmion number, have been observed both 
numerically and experimentally \cite{Leonov14,skyrmionium1,skyrmionium2,skyrmionium3,
skyrmionium4,skyrmionium5,skyrmionium6,skyrmionium7,skyrmionium8,skyrmionium9}. 
A skyrmionium can also be embedded inside another even larger skyrmion to form 
a target skyrmion of 3 layers with net skyrmion number 1. 
In fact, this cascade process can continue to form target skyrmions of any cascade
levels with skyrmion number 0 or 1 for even and odd numbers of layers, respectively. 
Target skyrmions cannot be continuously transformed into a single ferromagnetic 
domain structure or an elementary skyrmion. Also, so-called skyrmion bags, 
spin structures of arbitrary number of skyrmions inside another larger skyrmion, 
have also been observed \cite{Leonov14,skyrmionbag1,skyrmionbag2,skyrmionbag3,
skyrmionbag4,skyrmionbag5,skyrmionbag6}. Their skyrmion numbers can be arbitrary 
integers $Q$ and they cannot be continuously deformed into $Q$ spatially separated 
elementary skyrmions. Similar to the target skyrmions, one can put one or 
many skyrmion bags into another even bigger skyrmion to form a cascade skyrmion 
bag of 3 layers. The process can also continue to have various cascade skyrmion 
bags of arbitrary layers. Although composite skyrmions have been found in 
various simulations and experiments, there is no clear understanding of their 
existence conditions. They were found through trial and error or hunch so far.   

The increasing number of newly discovered topologically non-trivial 
localized spin structures in chiral magnetic films reminisces the 
discovery of a huge number of elementary particles in the first half of 
last century that resulted in the quark model of elementary particles. 
It calls for a ``quark model" to properly organize these newly discovered 
topological magnetic structures. In this paper, we consider a perpendicularly 
magnetized chiral magnetic film with the Dzyaloshinskii–Moriya interaction 
(DMI) and normal ferromagnetic exchange interaction characterized by parameter 
$D$ and $A$, respectively, as well as magnetic anisotropy measured by $K$. 
We show that the stability and properties of various types of composite skyrmions 
is very sensitive to a parameter defined as $\kappa\equiv (\pi D)^2/(16AK)$. 
Near $\kappa=1$ that separates isolated circular skyrmions from condensed stripe 
skyrmions \cite{paper2}, the skyrmions show strong particle-continuum-medium duality. 
In contrast to the general belief that a skyrmion is a fundamental object indivisible. 
A skyrmion behaves sometimes like a continuum medium that can be tore apart to embed 
one or more than one skyrmions or composited skyrmions although it often appears as 
a particle. In another word, one or many circular and stripe skyrmions can be hold 
or be glued by another larger skyrmion, which acts as a glue or a bag, to form a 
skyrmionium or skyrmion bag/cluster. The embedding can continue layer by 
layer to form all kinds of target skyrmions and cascade skyrmion bags. 

Specifically, following interesting results were obtained in the absence of the 
external magnetic field for $\kappa\leq 1$ where isolated circular skyrmions 
are metastable and $\kappa>1$ where condensed stripe skyrmions are stable. 
1) Topologies of stable static magnetic textures are fully determined by 
$\kappa$, and $A/D$ defines their length scale. 2) The maximal number of 
cascade layers in target skyrmions for $\kappa\leq 1$ increases monotonically 
from 1 at $\kappa=0.7$, to 2 at $\kappa=0.8$, and to 4 at $\kappa=0.94$, 
and eventually diverges at $\kappa=1$ for an infinite large system. 
3) An unstable target skyrmion or skyrmion bag for $\kappa< 1$ can be stabilized by 
inserting enough skyrmions inside the innermost bag and/or the next innermost bag. 
The maximal number of skyrmions needed to stabilize a skyrmion bag decreases from 
a larger value at small $\kappa$, say 4 at $\kappa=0.7$, to the lowest possible 
1 when $\kappa$ approaches 1. This number increases also as the number of cascade 
layers increases. Furthermore, skyrmion size inside the innermost bag increases 
with the number of skyrmions inside the bag, and approaches the size of isolated 
elementary skyrmions. 4) For $\kappa>1$, the number of target stripe skyrmions 
and cascade stripe skyrmion bags can be any number as long as the average space-size 
occupied by each skyrmion is larger than twice of elementary stripe width. 
Furthermore, stripe width and stripe spin profile of any target stripe skyrmions and 
cascade stripe skyrmion bags are the same as those of elementary stripe skyrmions. 
5) When the average skyrmion-skyrmion distance inside a stripe skyrmion bag 
is comparable to the elementary stripe width for $\kappa>1$, stripe 
skyrmions in the innermost layer become disk-like objects and form an SkX. 
The paper is organized as follows. The model and methodology is present in the 
next section. We prove there that stable/metastable spin structures are fully 
determined by $\kappa$ with the fundamental length scale of $4A/(\pi D)$.   
Section III is the results, and discussion and conclusion are given in Sec. IV.

\section{Model and Methodology}
To demonstrate particle-continuum-medium duality of skyrmions, we consider a 
thin chiral magnetic film of thickness $d$ in the xy-plane. The magnetic 
energy of a spin structure $\vec{m}$ is 
\begin{equation}
	\begin{aligned}
&E=d\iint \lbrace A|\nabla \vec{m}|^2+D[(\vec{m}\cdot\nabla)m_z-m_z\nabla
\cdot\vec{m}]+ \\&K_{\rm u}(1-m_z^2)-\frac{1}{2}\mu_0 M_{\rm s}\vec H_{\rm d}
\cdot\vec{m}+\mu_0 H M_{\rm s}(1-m_z)\rbrace\,\mathrm{d}S,
	\end{aligned}
	\label{energy}
\end{equation}
where $K_{\rm u}$, $H$, $M_{\rm s}$, $\vec{H}_{\rm d}$ and $\mu_0$ are the 
magneto-crystalline anisotropy, perpendicular magnetic field, the saturation 
magnetization, the demagnetizing field and the vacuum permeability, respectively. 
The energy of ferromagnetic state $m_z=1$ is chosen as the energy reference of $E=0$. 
For an ultra thin film, demagnetization effect can be 
included in the effective anisotropy $K=K_{\rm u}-\mu_0M_{\rm s}^2/2$. 
This is a good approximation when the film thickness $d$ is much smaller 
than the exchange length \cite{Xiansi}. 

One can recast Eq. \eqref{energy} to reveal how model parameters affect 
stable/metastable magnetic structures, 
\begin{equation}
\begin{aligned}
E &= d\int \int\lbrace A|\nabla\mathbf{m}|^2+D[m_z\nabla\cdot\mathbf{m}
-(\mathbf{m}\cdot\nabla)m_z]\\ &\quad
+K(1-m_z^2)+\mu_0 H M_{\rm s}(1-m_z)\rbrace \mathrm{d}x\mathrm{d}y\\
&= d\kappa K \int\int\lbrace |L\nabla\mathbf{m}|^2+
\frac{4L}{\pi}[m_z\nabla\cdot\mathbf{m}-(\mathbf{m}\cdot\nabla)m_z] \\	
&\quad  +\frac{1}{\kappa}(1-m_z^2)+\frac{2}{\kappa'}(1-m_z)  
\rbrace  \mathrm{d}x\mathrm{d}y\\
&= dA\int\int\lbrace |\nabla\mathbf{m}|^2+
\frac{4}{\pi}[m_z\nabla\cdot\mathbf{m}-(\mathbf{m}\cdot\nabla)m_z]
\\ &\quad	+\frac{1}{\kappa}(1-m_z^2) 
+\frac{2}{\kappa'}(1-m_z)\rbrace  \mathrm{d}x\mathrm{d}y, 
\end{aligned}	 
\label{energy1}	
\end{equation} 
where $\kappa=\frac{\pi^2 D^2}{16AK}$, $\kappa'=\frac{\pi^2D^2}{8A\mu_0M_sH}$, 
and $L=4A/(\pi D)$. The stable/metastable spin structures minimize the last 
integral of above equation \cite{paper4} where $x$ and $y$ is in the units of $L$. 
In the absence of external magnetic field ($1/\kappa'=0$) considered here, all 
stable/metastable magnetic structures satisfy following dimensionless equation, 
\begin{equation}
\nabla^2\vec{m}+\frac{4}{\pi}\left[(\nabla \cdot \vec{m})\hat{z}-\nabla 
m_z\right]+\frac{1}{\kappa}m_z\hat{z}=0. 
\label{sseq}
\end{equation} 
In another word, stable/metastable magnetic structures are fully determined by 
$\kappa$, not by $A$, $D$, $M_s$, and $K$ separately. $L$ is the length scale. 
This fact greatly simplifies our study of skyrmion static properties, and 
all stable magnetic structures in general. It allows us to compare skyrmions 
in different chiral magnets in terms of $\kappa$. Varying $\kappa$ over 
all possible values corresponds to studying all possible magnetic films.
Isolated circular skyrmions are metastable and their energies are $E=8\pi Ad
\sqrt{1-\kappa}$ when $H=0$ and $\kappa\leq 1$ \cite{Xiansi}. Obviously, the 
formation energy of an isolated skyrmion goes to zero as $\kappa$ approaches 1. 
Earlier studies \cite{paper1,paper2,paper3,paper4} showed that $\kappa=1$ 
separates isolated circular skyrmions from condensed stripe skyrmions.
The natural skyrmion morphologies are stripes in order to take the advantages 
of negative skyrmion formation energy when $\kappa>1$ \cite{paper3}. 

Spin dynamics in a magnetic field is governed by the Landau-Lifshitz-Gilbert 
(LLG) equation,
\begin{equation}
\frac{\partial \vec m}{\partial t} =-\gamma\vec m \times \vec H_{\rm eff} +
\alpha \vec m \times \frac{\partial \vec m}{\partial t}, 
\label{llg}
\end{equation}
where $\gamma$ and $\alpha$ are respectively gyromagnetic ratio and Gilbert 
damping constant. $\vec H_{\rm eff}=\frac{2A}{\mu_0M_{\rm s}} \nabla^2\vec m+
\frac{2K_{\rm u}}{\mu_0M_{\rm s}}m_z\hat z+H\hat z+\vec H_{\rm d}+\vec H_{\rm DM}+
\vec{h}$ is the effective field which includes the exchange field, the anisotropy field, 
the external magnetic field along $\hat z$, the demagnetizing field, the DMI field 
$\vec H_{\rm DM}$, and the thermal fluctuating field $\vec{h}$ (in the case of 
finite temperature), respectively \cite{MuMax3}. In the absence of an energy source, 
the steady state solutions of Eq. \eqref{llg} are stable/metastable spin textures. 
We will use MuMax3 \cite{MuMax3} to obtain all kinds of composite skyrmions.   
The initial configurations are important in this endeavour. According to our 
previous study \cite{paper1}, each nucleation domain in a chiral magnetic film, 
in which skyrmions are stable/metastable, develops into a skyrmion. 
Thus, various domains embedded inside other domains can develop into various 
composite skyrmions, as demonstrated below.

\begin{figure*}
\centering
\includegraphics[width=17cm]{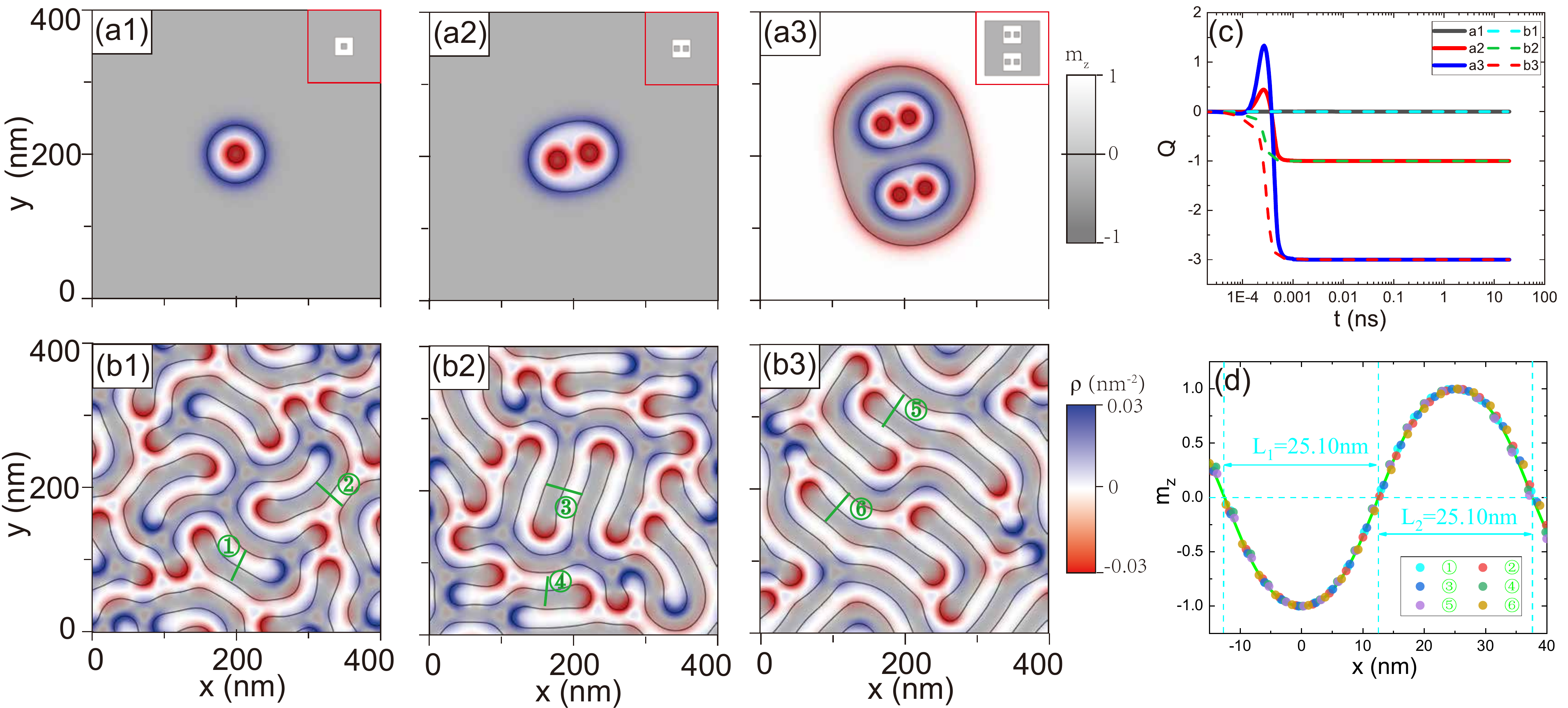}\\
\caption{Typical structures and profiles of composite circular skyrmions for 
$\kappa=0.9$ (a1)-(a3) and composite stripe skyrmions for $\kappa=4$ (b1)-(b3). 
One skyrmion inside another one (a1) and (b1) from the same initial configuration 
shown in the inset of (a1); two skyrmions inside another larger one (a2) and (b2) from 
the initial configuration shown in the inset of (a2); two composite skyrmions of (a2) 
or (b2) inside another larger skyrmion (a3) and (b3) from the initial configuration 
shown in the inset of (a2). (c) Time evolution of topological charges for structures 
(a1) (the black solid line); (a2) (the red solid line); (a3) (the blue solid line); 
(b1) (the cyan dash line); (b2) (the green dash line); and (b3) (the red dash line).
$t$ is in the logarithmic scale. Skyrmion charge density $\rho$ is encoded by colours 
(the blue for positive and the red for negative) while the gray-scale encodes $m_z$. 
(d) Spin profiles $m_z(x)$ along the green lines labelled by \textcircled{n} in (b1)-(b3). 
The symbols are numerical data and the solid lines are the theoretical fits to 
$\cos\Theta (x)=-\frac{\sinh^2(L/2w)
-\sinh^2(x/w)}{\sinh^2(L/2w)+\sinh^2(x/w)}$ for $-L/2<x<L/2$ and 
$\cos\Theta (x)=\frac{\sinh^2(L/2w)-\sinh^2\left[(x-L)/w\right]}{\sinh^2(L/2w)
+\sinh^2\left[(x-L)/w\right]}$ for $L/2<x<3L/2$ with $L=25.10$ nm 
and $w=7.63$nm.}
\label{fig1}
\end{figure*}

\section{Results}

\subsection{Target skyrmions and cascade skyrmion bags at $\kappa=0.9$ and $\kappa=4$}

Stable/metastable composite skyrmions are expected to more favourably appear 
in chiral magnetic films with $\kappa$ slightly smaller than 1 or $\kappa>1$ 
because creating an individual skyrmion gains energy for $\kappa>1$ and costs 
nearly zero energy near $\kappa=1^-$. To demonstrate that this expectation is 
indeed true, we choose $A=4\, \rm{pJ\,m^{-1}}$, $D=1\, \rm mJ\,m^{-2}$, and $M_s=
0.15\, \rm MA\,m^{-1}$, values in the range of $A=2\sim 16\, \rm pJ\, \rm m^{-1}$, 
$D=0.68\sim 4\, \rm mJ\, \rm m^{-2}$, $K_u=0.2\sim 2.5\, \rm MJ\, \rm m^{-3}$ 
and $M_s=0.65 \sim 1.1 \, \rm MA\, \rm m^{-1}$ for chiral magnets of $\rm Ir/ 
\rm Fe/ \rm Co/ \rm Pt$ multilayer and $\rm PdFe/ \rm Ir$ bilayers \cite
{Romming,PdFeIr,size2015,multilayer}. $K_u$ is used to simulate samples with 
different $\kappa$ since in reality $K_u$ is sensitive to chemical compositions, 
fabrication process, structures, and the temperature. $K_u=0.185\, \rm MJ\,m^{-3}$ 
and $K_u=0.0526\, \rm MJ\,m^{-3}$ give $\kappa=0.9$ and $\kappa=4$, respectively. 
As explained in the previous section, the stable/metastable structures for  
the same $\kappa$ are invariant when they are scaled by $4A/(\pi D)$. 
Figure \ref{fig1} shows the typical structures of composite skyrmions 
in a sample of $400\,$nm $\times400\,$nm $\times0.5\,$nm in size 
with periodic boundary conditions along both $x$ and $y$ directions. 
Throughout this study, skyrmion structures are encoded in the gray-scale 
for $m_z$ and by colors for skyrmion charge density $\rho$ defined as $\rho
=\mathbf m\cdot(\partial_x\mathbf m\times\partial_y\mathbf m)/(4\pi)$.

Figure \ref{fig1}(a1) is a stable skyrmionium, starting from an initial 
configuration of a $30\, \rm nm \times 30\, \rm nm \times 0.5\, \rm nm$ 
$(m_z=-1)$-domain inside another $(m_z=1)$-domain of $100\, \rm 
nm \times 100\, \rm nm \times 0.5\, \rm nm$ at the center in the 
background of $m_z=-1$ as shown in the inset (the red-line square). 
It forms shortly a stable skyrmionium of skyrmion number (charge) 
0 with an inner core of topological charge -1 and an outer shell of 
topological charge 1, separated by the white-color circle on which $\rho=0$. 
The topological charge remains 0 shortly after $10\,$ps as shown by the black 
solid curve in Fig. \ref{fig1}(c). Figure \ref{fig1}(a2) is a stable composite 
topological skyrmion with two skyrmions inside another larger skyrmion, starting 
from an initial configuration (shown in the inset) of two $30$ nm $\times$ $30\,
\rm nm \times 0.5\,\rm nm$ $(m_z=-1)$-domains inside another $(m_z=1)$-domain of 
$100\,\rm nm\times 100\, \rm nm\times 0.5\,\rm nm$ in the background of $m_z=-1$. 
Two skyrmions enclosed by the white-color contour of $\rho=0$ have skyrmion 
number -1 each while the total charge outside of the contour is 1. 
The red solid curve in Fig. \ref{fig1}(c) is the time evolution of the total 
topological charge $Q(t)$, $Q(t)=-1$ shortly after $10\,$ps evolution. Similarly, 
Fig. \ref{fig1}(a3) is a stable cascade skyrmion bag of 3 cascade layers: two 
composite skyrmions shown in Fig. \ref{fig1}(a2) inside another larger skyrmion. 
This cascade skyrmion bag comes from an initial configuration consisting of two 
composite domains as that in the inset of Fig. \ref{fig1}(a2) inside another 
larger domain. As expected, this composite skyrmion has a skyrmion number of $4
\times (-1)+2\times 1+(-1)=-3$ as shown by the blue curve in Fig. \ref{fig1}(c).
Interestingly, $Q(t)$ exhibits a peak with a positive value larger than 1 before 
reaching its final negative value and its cause is unknown yet. 

In order to investigate whether the above results are also true for stripe 
skyrmions when skyrmion formation energy is negative, we repeat the above 
simulations for film with $\kappa=4$. Figures \ref{fig1}(b1)-(b3) show three 
stable structures of composite stripe skyrmions: stripe skyrmionium of 
topological charge 0 or one stripe skyrmion inside another larger stripe 
skyrmion (b1); a composite stripe skyrmion bag of topological charge -1 with two 
stripe skyrmions inside another larger stripe skyrmions (b2); a composite cascade 
stripe skyrmion bag of topological charge -3 with two composite stripe skyrmion 
bags as shown in Fig. \ref{fig1}(b2) inside another stripe skyrmions (b3). 
As a direct comparison to the case of $\kappa=0.9$, the initial configurations are 
exactly the same as their corresponding counterparts in Figs. \ref{fig1}(a1)-(a3). 
The dash curves in Fig. \ref{fig1}(c) are the time evolution of topological 
charge of different structures. Different from the case of $\kappa<1$, 
$Q(t)$ reaches monotonically to its final stable values. Also, different 
from the case of $\kappa<1$ where the skyrmion sizes in different layers 
are different, the stripe width in the current case are the same. 
Even more interestingly, spin profile are the same as that for elementary stripe 
skyrmion \cite{paper2}, $\Theta(x)=2\arctan[\frac{\sinh(L/w)}{\sinh(x/w)}]$ 
for $m_z>0$ and $\Theta(x)=2\arctan[\frac{\sinh(x/w)}{\sinh(L/w)}]$ for 
$m_z<0$, respectively, with $|x| \leq L/2$. $\Theta$ is the polar angle of the 
magnetization at $x$ and $x = 0$ is the center of a stripe where $m_z =\pm 1$. 
The symbols in Fig. \ref{fig1}(d) are numerical data along the green line labelled 
by green \textcircled{n} in Figs. \ref{fig1}(b1)-(b3). The solid curve is the fit 
of spin profiles to $\cos\Theta (x)=-\frac{\sinh^2(L/2w)-\sinh^2(x/w)}{\sinh^2
(L/2w)+\sinh^2(x/w)}$ for $-L/2<x<L/2$ and $\cos\Theta (x)=\frac{\sinh^2(L/2w)-
\sinh^2\left[(x-L)/w\right]}{\sinh^2(L/2w)+\sinh^2\left[(x-L)/w\right]}$ for 
$L/2<x<3L/2$ with $L=25.10\,$nm and $w=7.63\,$nm. All data from different 
stripes falling onto the same curve demonstrates that stripes, building blocks 
of the pattern, are identical. 

\subsection{Maximal layer number of target skyrmions for $\kappa\leq 1$}
\begin{table}[htbp]
\setlength{\tabcolsep}{2mm}{
\caption{The value of material parameters $K$, $K_u$ and $\kappa$ for 
8 different films.		}
\label{table1}	
\begin{tabular}{llll}
\hline\hline\noalign{\smallskip}
Films &$K_u (\, \rm MJ\, m^{-3}\, )$
&$K (\, \rm MJ \,  m^{-3}\, )$ &$\kappa$
\\ \noalign{\smallskip}\hline\noalign{\smallskip}
1 & 0.2344  & 0.2203 & 0.7        \\ 
2 & 0.2197  & 0.2056 & 0.75        \\ 
3 & 0.2068  & 0.1927 & 0.8       \\ 
4 & 0.1955  & 0.1814 & 0.85       \\
5 & 0.1854  & 0.1713 & 0.9        \\
6 & 0.1817  & 0.1676 & 0.92       \\
7 & 0.1781  & 0.1640 & 0.94        \\
8 & 0.1747  & 0.1606 & 0.96        \\
\noalign{\smallskip}\hline\hline
\end{tabular} } 	
\end{table}

\begin{figure*}[htbp]
\centering
\includegraphics[width=17cm]{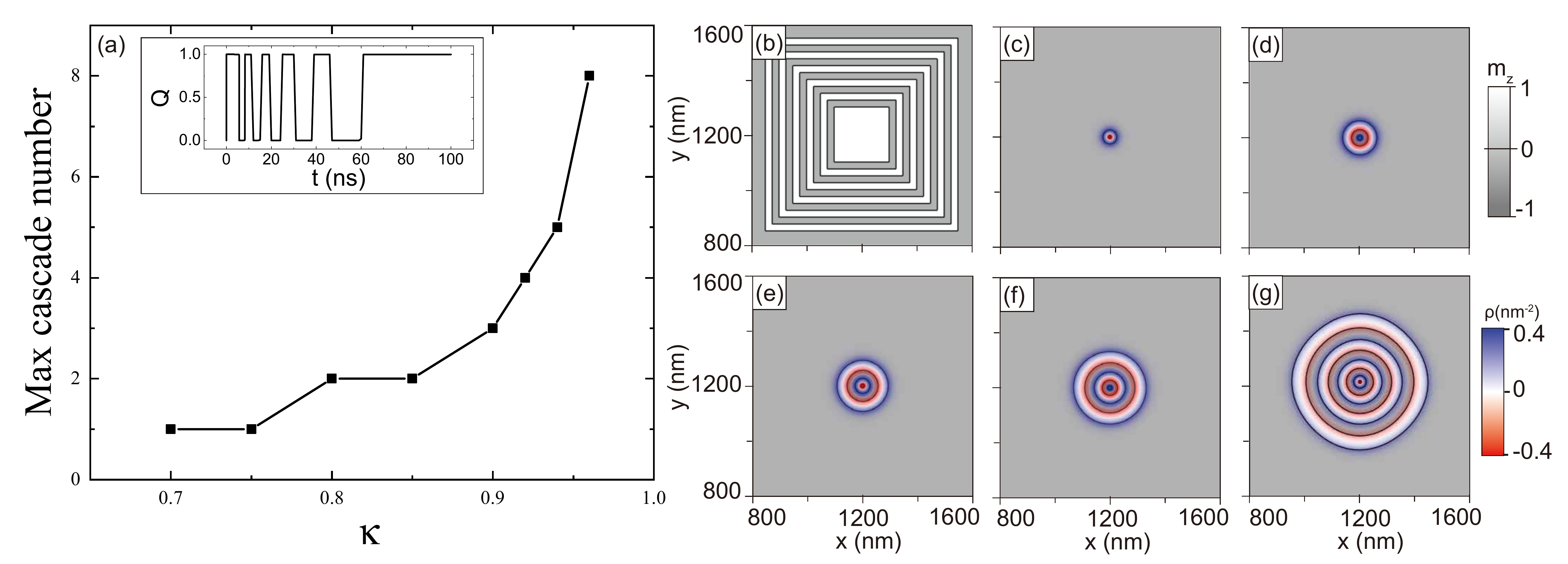}
\caption{(a) $\kappa$-dependence of the maximal number of layers of target skyrmions. 
The inset shows how skyrmion number of unstable composite spin structure varies with time 
$t$ when $\kappa=0.7$. The final stable spin texture is an elementary skyrmion of $Q=1$. 
(b) The initial configuration of a cascaded domain-inside-domain structure of $n=11$. 
(c)-(g) The final metastable target skyrmions from the initial configuration of 
(b) for various $\kappa$: Skyrmionium for $\kappa = 0.8$ (c) and target skyrmions 
of 3 layers for $\kappa=0.9$ (d), of 4 layers for $\kappa=0.92$ (e), of 5 
layers for $\kappa=0.94$ (f), and of 8 layers for $\kappa=0.96$ (f).
Color bar denotes skyrmion charge density and the gray-scale encodes $m_z$. } 
\label{fig2}
\end{figure*}
\begin{figure*}[htbp]
\centering
\includegraphics[width=17cm]{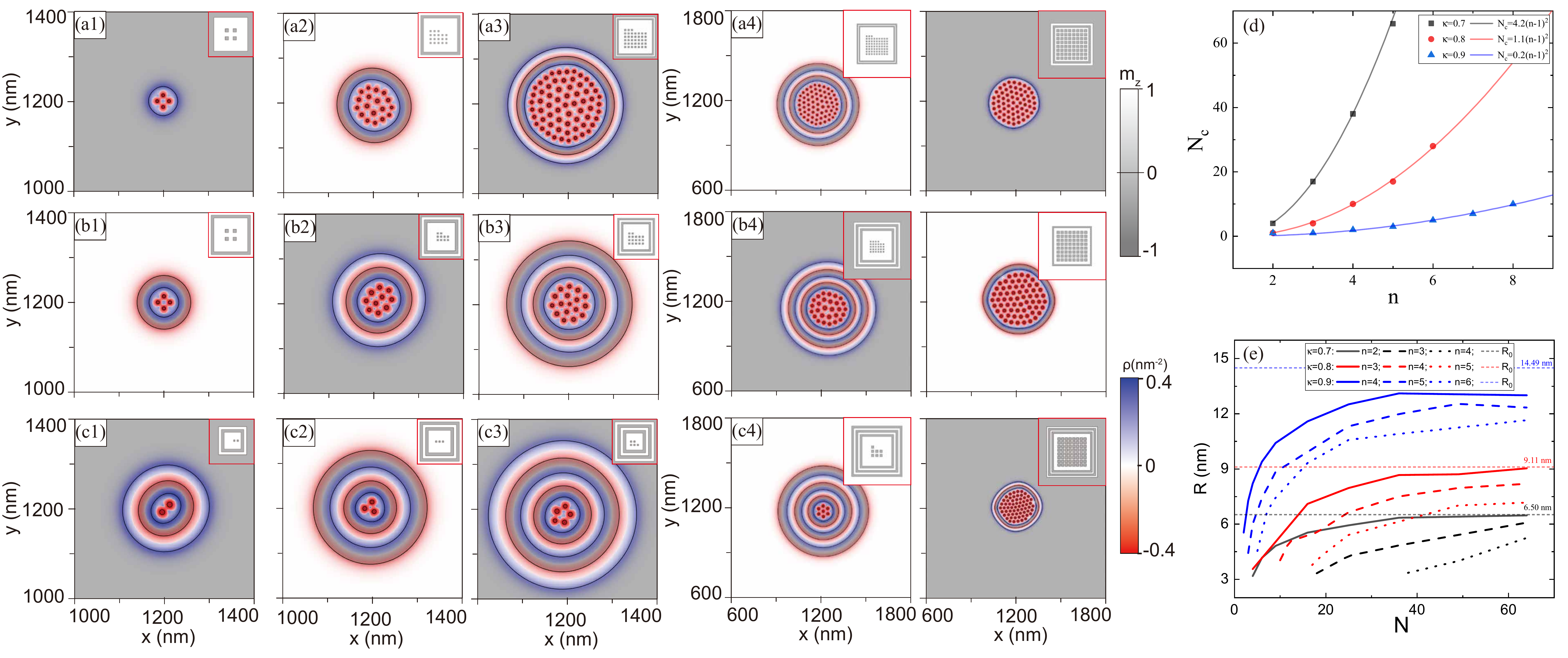}
\caption{Various stable cascade skyrmion bags of $n=2, 3, 4,$ and 5 cascade 
layers for $\kappa=0.7, 0.8,$ and 0.9. (a1)-(a4) Stable cascade skyrmion bags 
of $n=2,$ 3, 4, and 5 layers with minimal number of $N_c=4$ (a1), 18 (a2), 38 
(a3), and 66 (a4) skyrmions in the innermost bag for $\kappa=0.7$. 
(a5) A stable skyrmion bag of $N=64\gg N_c$ and $n=2$ for $\kappa=0.7$. 
(b1)-(b4) Stable cascade skyrmion bags of $n=3$, 4, 5, and 6 layers with minimal 
number of $N_c=4$ (b1), 10 (b2), 17 (b3), and 28 (b4) skyrmions in the innermost 
bag for $\kappa=0.8$. (b5) A stable cascade skyrmion bag of $N=64\gg N_c$ and 
$n=3$ for $\kappa=0.8$. (c1)-(c4) Stable cascade skyrmion bags of $n=4,$ 5, 6 and 
7 layers with minimal number of	$N_c=2$ (c1), 3 (c2), 5 (c3), and 7 (c4) in the 
innermost bag for $\kappa=0.9$. The insets are the initial configurations. 
(c5) A stable cascade skyrmion bag of $n=4$ and $N=64\gg N_c$ skyrmions in the 
innermost bag for $\kappa=0.9$. (d) $N_c$ vs. n. Dots are numerical data for 
$\kappa=0.7$ (the black); 0.8 (the red) and 0.9 (the blue). The solid lines are 
the fits of data to $N_c=a(\kappa)(n-1)^2$, with $a(0.7)=4.2$, $a(0.8)=1.1$, and 
$a(0.9)=0.2$. (e) Skyrmion size in the innermost bag as a function of the number 
of skyrmions $N$ in the innermost bag for $\kappa=0.7$ and $n=2$ (the black full 
line); $\kappa=0.7$ and $n=3$ (the black dash line; $\kappa=0.7$ and $n=4$ (the 
black point line); $\kappa=0.8$ and $n=3$ (the red full line); $\kappa=0.8$ and 
$n=4$ (the red dash line); $\kappa=0.8$ and $n=5$ (the red point line); $\kappa=0.9$ 
and $n=4$ (the blue full line); $\kappa=0.9$ and $n=5$ (the blue dash line); 
$\kappa=0.9$ and $n=6$ (the blue point line) and elementary skyrmion size for 
$\kappa=0.7$ (the black short dash line); $\kappa=0.8$ (the red short dash line); 
$\kappa=0.9$ (the blue short dash line). Color bar denotes skyrmion charge density 
and the gray-scale encodes $m_z$.}
\label{fig3}
\end{figure*}

When $\kappa<1$, the skyrmion formation energy of an elementary skyrmion is 
$E=8\pi A \sqrt{1-\kappa}$ \cite{Xiansi}, and skyrmion size and skyrmion wall 
width are $R=\pi D/ (4K\sqrt{1-\kappa})$ and $w=\pi D/ (4K)$, respectively. 
The energy cost is mainly from the skyrmion wall region. The size and 
wall regions in a composite skyrmion is obviously larger than that in 
an elementary skyrmion, resulting in an increase of energy-cost.
Thus, stability of various composite skyrmions should be sensitive to how far 
$\kappa$ is away from 1 at which the energy cost for creating a wall is zero. 
Although an isolated skyrmion is expected to be metastable when $\kappa\leq 1$ 
as long as the skyrmion size is much bigger than the lattice constant $a$, 
$R>a$ or $\kappa >1-[\pi D/ (4Ka)]^2$, in order to validate the continuum 
description of the films by Eqs. \eqref{energy}, \eqref{energy1}, and 
\eqref{sseq}, the maximal number of cascade layers of target skyrmions should 
change with $\kappa$. In order to test this hypothesis, we consider 8 samples 
with various $K_u$ listed in Tab. \ref{table1} while all other model 
parameters are unchanged. The corresponding $\kappa$ values are 0.7, 0.75, 
0.8, 0.9 0.92, 0.94 and 0.96. We started with a cascade domain-inside-domain 
structure as shown in Fig. \ref{fig2}(b), consisting of 6 white $(m_z=1)$-domains 
and 6 gray $(m_z=-1)$-domains. They were segregated by 11 concentric squares 
of side lengths of $750,\, 700,\, 650,\, 600,\, 550,\, 500,\, 450,\, 400,\, 350,
\,300,\,250\,$nm at the film center. The sample size is of $2400\,\rm nm \times 
2400\, \rm nm \times 0.5\, \rm nm$, large enough so that the final stable spin 
structures do not depend on the sample size. The mesh size is $1\,$nm $\times 
1\,$nm $\times 0.5\,$nm in the simulations. For $\kappa=0.7$, the initial 
configuration quickly evolves into an unstable shell structure of skyrmions 
number $Q=1$ in less than $10\,$ps as shown in the inset of Fig. \ref{fig2}(a) 
in the first jump around $t=0$ from $Q=0$ to 1 in $Q(t)$-$t$ curve. 
Simulations show that the shell structure shrinks its size, and the 
innermost spin structure disappears one by one from the center. 
As revealed in $Q(t)$-$t$ curve, $Q$ jumps back and forth between 1 and 0 
each time when one shell disappears. The final metastable spin 
structure is an elementary skyrmion of $Q=1$, indicating that 
target skyrmions and skyrmionium are not stable when $\kappa=0.7$. 

For $\kappa=0.8$, the final stable spin structure is a skyrmionium as shown 
in Fig. \ref{fig2}(c). Figures \ref{fig2}(d)-(g) are the final stable target 
skyrmions for $\kappa=0.9$, 0.92, 0.94, and 0.96 with maximal layer number 
3 (d), 4 (e), 5 (f), and 8 (g), respectively. The $\kappa$-dependence of 
the maximal number of cascade layers in the final stable/metastable target 
skyrmions is plotted in Fig. \ref{fig2}(a), showing the number increases 
to a very large value as $\kappa$ approaches 1. 

\subsection{Stabilization and properties of cascade skyrmion bags for $\kappa\leq 1$}

\begin{figure*}[htbp]
\centering
\includegraphics[width=17cm]{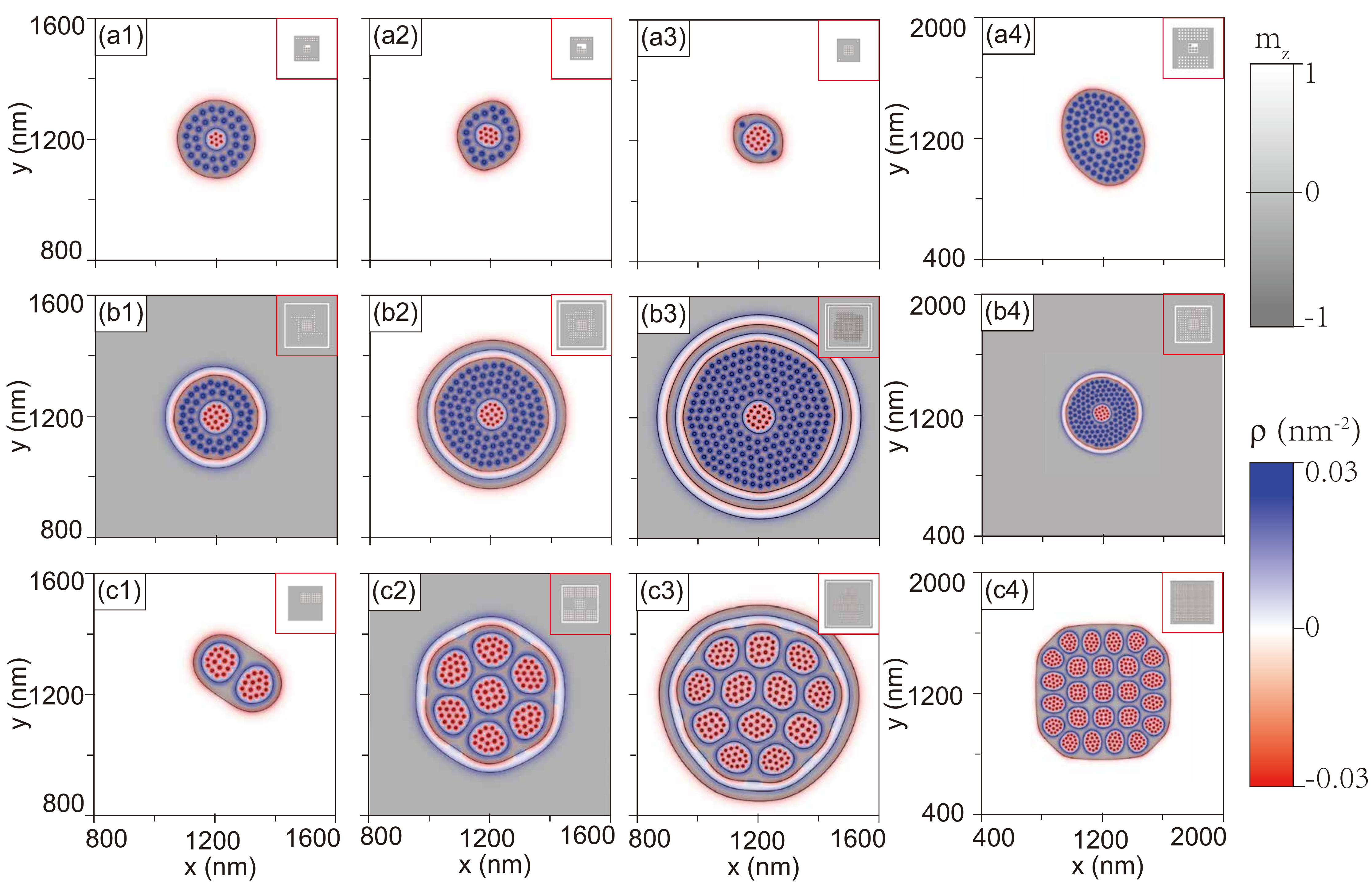}
\caption{Various composite skyrmion of $\kappa=0.7$. (a1)-(a3) Stable cascade 
skyrmions of $N=7$ and minimal $N_{1,c}=32$ (a1); $N=10$ and $N_{1,c}=18$ 
(a2); $N=16$ and $N_{1,c}=2$ (a3). (a4) A stable cascade skyrmion 
of $N=7$ and $N_1=80$ which is much larger than the critical value of 32.
(b1)-(b3) Stable cascade skyrmion bags of $n=4,$ 5, and 6 layers and $N=16$ 
and $N_{1,c}=44, 104,$ and 200. (b4) A stable cascade skyrmions of $n=4$, 
$N=10$, and $N_1=128$ which is much larger than the critical value of 44. 
(c1)-(c3) Four stable skyrmion superstructures of $n=3,$ 4, and 5 layers 
with minimal 2 (c1), 7 (c2), 12 (c3) skyrmion bags in the innermost layers. 
Each skyrmion bag contains $N=16$ elementary skyrmions. (c4) A stable 
skyrmion superstructure of 3 layers with 25 skyrmion bags of $N=16$.
Color bar denotes skyrmion charge density and the gray-scale encodes $m_z$.}
\label{fig4}
\end{figure*}
Figure \ref{fig2} shows the maximal number of layers in stable target skyrmions 
being 1 for $\kappa=0.7$, i.e. even a skyrmionium of $Q=0$ is not stable.
The energy of a skyrmion bag with many skyrmions in the innermost layer 
is higher than that of a skyrmionium. Thus, one may expect that a spin 
bag is more unstable than a skyrmionium for $\kappa=0.7$. Surprisingly and 
counter-intuitively, skyrmion bags with $N$ skyrmions inside another larger 
skyrmion are stable as long as $N$ is larger than a critical value of $N_c=4$ as 
shown in Fig. \ref{fig3}(a1) for $N=N_c=4$ and in Fig. \ref{fig3}(a5) for $N=64$. 
Their total net topological charges are respectively $Q=-3$ and $Q=-63$ obtained from 
the initial configurations of 4 and 64 $(m_z=-1)$-domains inside a $(m_z=1)$-domain 
as shown in the insets. Similarly, the cascade skyrmion bags of $n$ layers can also 
be stabilized by putting $N$ skyrmions in the innermost bag when $N$ is larger than 
a critical value $N_c(n)$ which depends on layer number of cascade skyrmion bag. 
Figures \ref{fig3}(a2)-(a4) are the stable cascade skyrmion bags of $n=3,$ 4 
and 5 with corresponding $N_c=18$, 38 and 66 when $\kappa=0.7$ respectively. 
These cascade skyrmion bags are obtained from the initial configurations shown 
in the insets. The net topological charges of these cascade skyrmion bags are 
$Q=-18$, -37 and -66. For $\kappa=0.8$, Figs. \ref{fig3}(b1)-(b4) show cascade 
skyrmion bags of $n=3$, 4, 5 and 6 with corresponding $N_c=4$, 10, 17 and 28. 
Figure \ref{fig3}(b5) shows a stable cascade skyrmion of 3 layers with $N=64$ 
which is larger than $N_c=4$. For $\kappa=0.9$, Figs. \ref{fig3}(c1)-(c4) show 
cascade skyrmion bags of $n=4$, 5, 6 and 7 with corresponding $N_c=2$, 3, 5 and 7. 
Figure \ref{fig3}(c5) shows a stable cascade skyrmion of 3 layers with 
$N=64$ which is larger than $N_c=2$.

Interestingly, we find that $N_c$ is proportion to $(n-1)^2$ for all our $\kappa$'s.
Figure \ref{fig3}(d) shows the relation between $N_c$ and $n$ for $\kappa=0.7$
(black dots); 0.8(red dots) and 0.9 (blue dots). Numerical data (symbols) fit 
well to $N_c=4.2(n-1)^2$ (the black line); $N_c=1.1(n-1)^2$ (the red line); 
$N_c=0.2(n-1)^2$ (the blue line). More interestingly, the skyrmion size $R$
of skyrmions in the innermost bag is smaller than that of elementary skyrmions, 
$R_0=\pi D/(4K\sqrt{1-\kappa})$, but it increases with $N$ as shown in Fig. 
\ref{fig3}(e).

In order to explain why a cascade skyrmion bag at a lower $\kappa$, say $\kappa
=0.7$, needs a minimal number of skyrmions in the innermost bag to stabilize 
itself, one may suspect that innermost bag size needs to be large enough so 
that geometrical restriction would not annihilates the skyrmions in the bag. 
If this is correct, one would expect that adding more skyrmions to the 
next innermost bag tends to destabilize a cascade skyrmion bag because 
skyrmions in the next innermost bag tends to reduce innermost bag size. 
And minimal number of skyrmions in the innermost bag of a stable cascade 
skyrmions increases with number of skyrmions in the next innermost bag. 
Strangely, this seemingly reasonable conjecture turns out to be incorrect. 
Contrary to this conjecture, we find the opposite results. As shown in Fig. 
\ref{fig4}(a1) for a stable cascade skyrmion bag of 3 layers ($n=3$) at 
$\kappa=0.7$, the cascade skyrmion bag have 7 skyrmions in the innermost bag and 
32 skyrmions in the next innermost bag. In Fig. \ref{fig3}(a2) we have seen that 
the minimal number of skyrmions in the innermost bag required to stabilize the 
cascade skyrmion is $N_c=18$ if no extra skyrmions are in the next innermost bag. 
In fact, all cascade skyrmion bags with 7 skyrmions in the innermost bag are stable 
as long as the number of skyrmions in the next innermost bag is larger than 32. 
Figure \ref{fig4}(a4) shows a cascade skyrmion bag of 3 layers with 7 
skyrmions in the innermost bag and 80 skyrmions in the next innermost bag.
The minimal number of skyrmions in the next innermost bag decreases as the 
number of skyrmions in the innermost bag increases. Equivalently, the minimal 
number of skyrmions in the innermost bag decreases with the increase of the 
number of skyrmions in the next innermost bag. Figures \ref{fig4}(a2)-(a3) are 
two stable cascade skyrmion bags of 3 layers with 10 and 16 
skyrmions in the innermost bag and corresponding minimal 18 and 2 
skyrmions in the next innermost bag, respectively. 

Above results are very sensitive to layer number $n$ in cascade skyrmion bags. 
If one keeps the number of skyrmions in the innermost bag of a cascade skyrmion 
bag below its minimal value such that the composite skyrmion by itself is unstable, 
the minimal number of skyrmions in the next innermost bag $N_{1,c}$(n) required 
to stabilize the cascade skyrmion bag increases with the cascade level. 
Figures \ref{fig4}(b1)-(b3) are stable cascade skyrmion bags of 4, 5, and 6 
and layers with 16 skyrmions in the innermost bag and minimal 44, 104 and 200 
skyrmions in the next innermost bag, respectively. Figure \ref{fig4}(b4) is 
a stable cascade skyrmion of 4 layers of $N=16$ and $N_1=128$ 
which is much larger than the corresponding critical value $N_{1,c}=44$. 

Instead of inserting elementary skyrmions into one particular cascade layer, 
one can also put skyrmion bags into it to obtain a more complicated skyrmion 
superstructure. Figures \ref{fig4}(c1)-(c3) are three such stable super skyrmion 
bags for $\kappa=0.7$ where skyrmioniun and other target skyrmions are unstable.
Different from Fig. \ref{fig3}(a1) where adding 4 elementary skyrmions into 
an unstable skyrmionium can obtain a stable skyrmion bag, Fig. \ref{fig4}(c1) 
is a stable skyrmion superstructure when we insert two skyrmion bags of $N=16$ 
into an unstable skyrmionium. Of course, this skyrmion superstructure has 
$n=3$ cascade layers and the total skyrmion number is $2\times (16-1)+1=31$.
2 is the minimal number of bags needed to stabilise the structure. 
All skyrmion superstructures with more than 2 skyrmion bags in an unstable skyrmionium 
are stable, as shown in Fig. \ref{fig4}(c4) with 25 skyrmion bags of $N=16$. 
The minimal number of bags required to stabilize an unstable target skyrmion of 
more layers increases with the target skyrmion layers. Figures \ref{fig4}(c2) and (c3) 
are the final stable skyrmion superstructures when minimal 7 and 12 skyrmion bag of 
$N=16$ are inserted into a target skyrmion of 3 (c2) and 4 (c3) layers, respectively. 
One notes that different sample sizes are used for these two cases in order to 
make sure that the results are not affected by the boundaries. 

\subsection{Resilience of stripe width of composite skyrmions for $\kappa>1$}
\begin{figure*}[htbp]
\centering
\includegraphics[width=17cm]{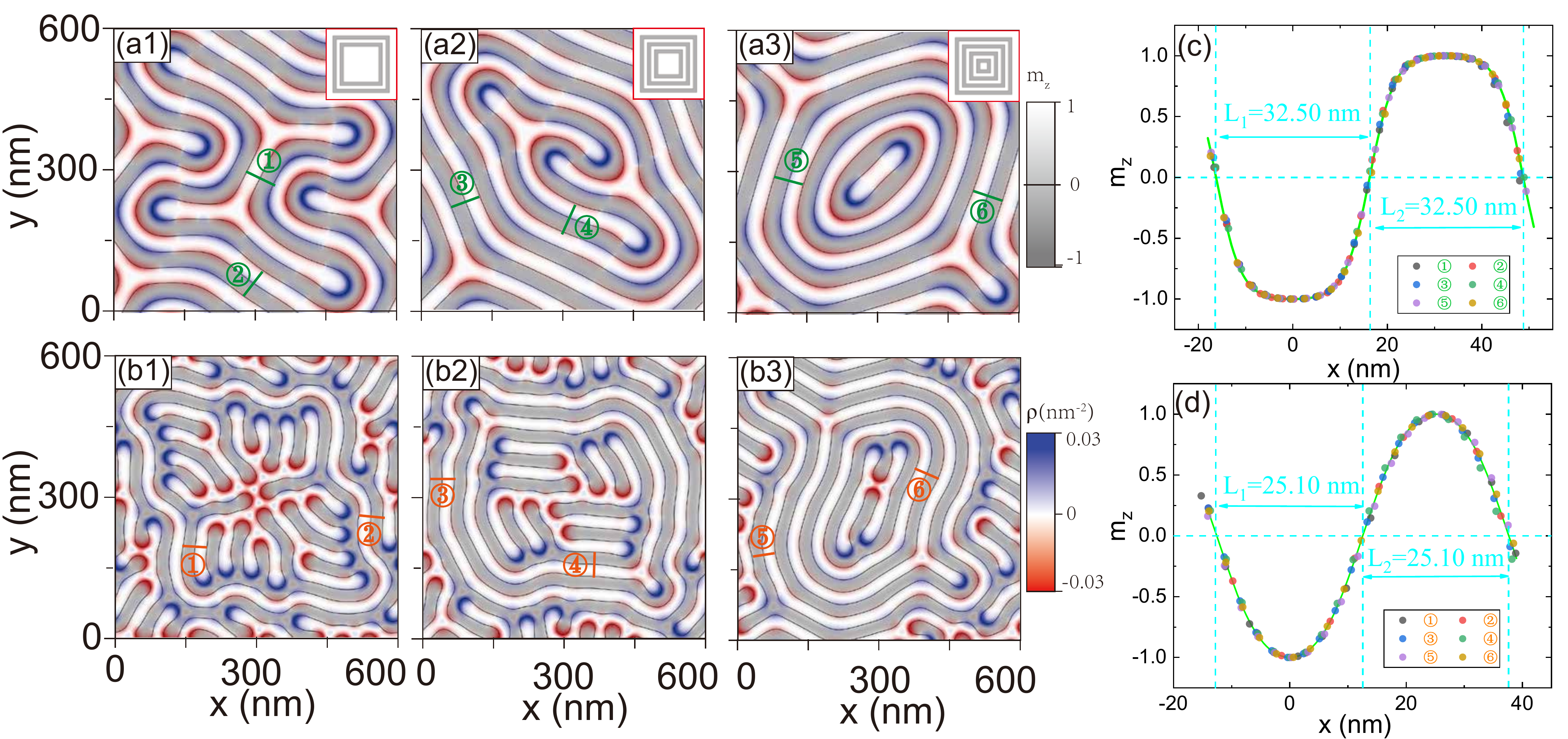}
\caption{Stable target stripe skyrmions of 4, 6, and 8 layers   
for $A/D= 4\,$nm and $\kappa=1.1$ (a1)-(a3) and $\kappa=4$ (b1)-(b3),
respectively. The sample sizes are $600\,\rm nm\times600\,\rm nm \times0.5
\,nm$ and the average layer width is larger than the elementary strip width. 
(a1) and (b1), (a2) and (b2), and (a3) and (b3) start respectively from the 
same initial configurations shown in the inset of (a1) (a2), and (a3). 
(c)-(d) Spin profile of $m_z(x)$ along the green lines and orange line
labelled by \textcircled{n} in (a1)-(b3). The symbols are numerical data and the 
solid lines are the theoretical fits to $\cos\Theta (x)=-\frac{\sinh^2(L/2w)
-\sinh^2(x/w)}{\sinh^2(L/2w)+\sinh^2(x/w)}$ for $-L/2<x<L/2$ and 
$\cos\Theta (x)=\frac{\sinh^2(L/2w)-\sinh^2\left[(x-L)/w\right]}{\sinh^2(L/2w)
+\sinh^2\left[(x-L)/w\right]}$ for $L/2<x<3L/2$ with $L=32.50\,$nm and 
$w=5.23\,$nm (c); and $L=25.10\,$nm and $w=7.63\,$nm (d). 
Color bar denotes skyrmion charge density and the gray-scale encodes $m_z$. }
\label{fig5}
\end{figure*}

\begin{figure*}[htbp]
\centering
\includegraphics[width=17cm]{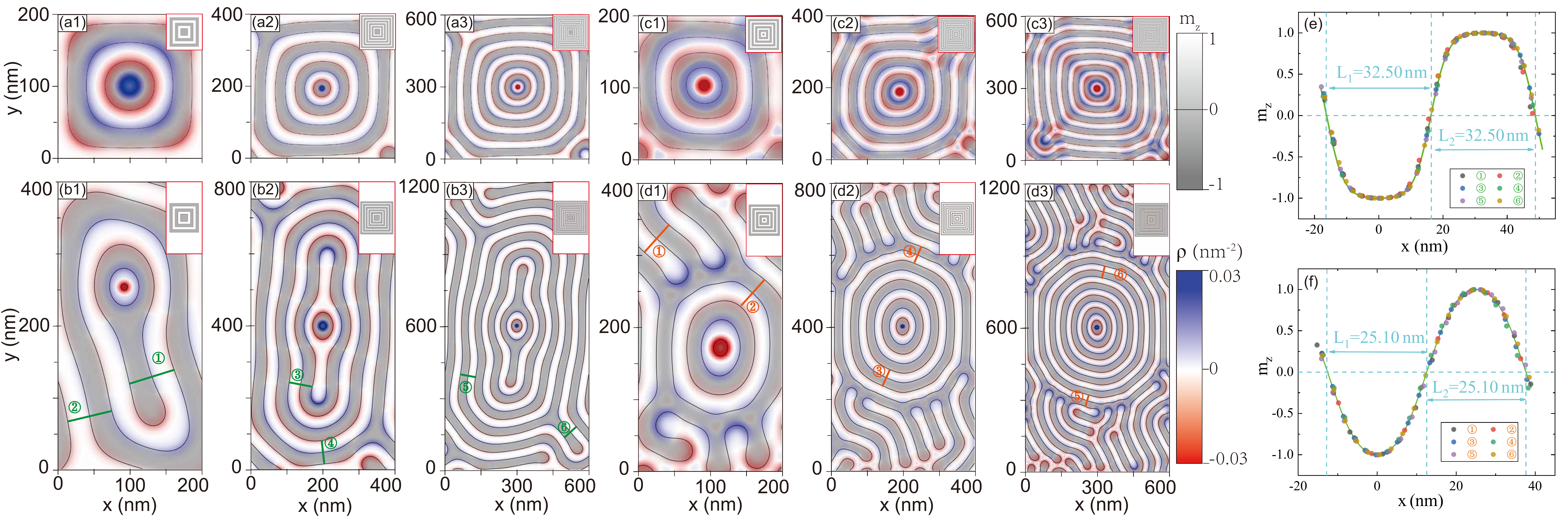}
\caption{Target stripe skyrmions with the maximal numbers of cascade layers 
$n_{max}$ for $\kappa=1.1$ and 4. The maximal numbers increase with sample sizes and 
are slightly larger than the ratio of sample size and the elementary stripe width. 
(a1)-(a3) Stable target skyrmions of $n_{max}=4$, 8 and 11 in $200\,\rm nm\times 200\,
\rm nm \times 0.5\,nm$, $400\,\rm nm\times 400\,\rm nm \times 0.5\,nm$ and $600\,\rm 
nm\times 600\,\rm nm \times 0.5\,nm$ samples for $\kappa=1.1$. (b1)-(b3) Stable target 
skyrmions of $n_{max}=5$, 8 and 12 in $200\,\rm nm\times 400\,\rm nm \times 0.5\,nm$, 
$400\,\rm nm\times 800\,\rm nm\times 0.5\,nm$ and $600\,\rm nm\times 1200\,\rm nm$ 
samples for $\kappa=1.1$. (c1)-(c3) Stable target skyrmions of $n_{max}=5$, 9 and 13 in 
$200\,\rm nm\times 200\,\rm nm \times 0.5\,nm$, $400\,\rm nm\times 400\,\rm nm \times 
0.5\,nm$ and $600\,\rm nm\times 600\,\rm nm \times 0.5\,nm$ samples for $\kappa=4$. 
(d1)-(d3) Stable target skyrmions of $n_{max}=5$, 10 and 14 in 
$200\,\rm nm\times 400\,\rm nm \times 0.5\,nm$, $400\,\rm nm\times 800\,\rm nm\times 
0.5\,nm$ and $600\,\rm nm\times 1200\,\rm nm$ for $\kappa=4$. Insets of (a1)-(d3) 
are initial configurations. (e)-(f) Spin profiles of stripes. The symbols are 
numerical data and the solid lines are the fits to $\cos\Theta(x)=-\frac{\sinh^2
(L/2w)-\sinh^2(x/w)}{\sinh^2(L/2w)+\sinh^2(x/w)}$ for $-L/2<x<L/2$ and $\cos\Theta 
(x)=\frac{\sinh^2(L/2w)-\sinh^2\left[(x-L)/w\right]}{\sinh^2(L/2w)+\sinh^2\left[
(x-L)/w\right]}$ for $L/2<x<3L/2$ with $L=32.50\,$nm and $w=5.23\,$nm (e); and
$L=25.10\,$nm and $w=7.63\,$nm (f). Color bar denotes skyrmion charge density and 
the gray-scale encodes $m_z$.}
\label{fig6}
\end{figure*}

\begin{figure*}[htbp]
\centering
\includegraphics[width=17cm]{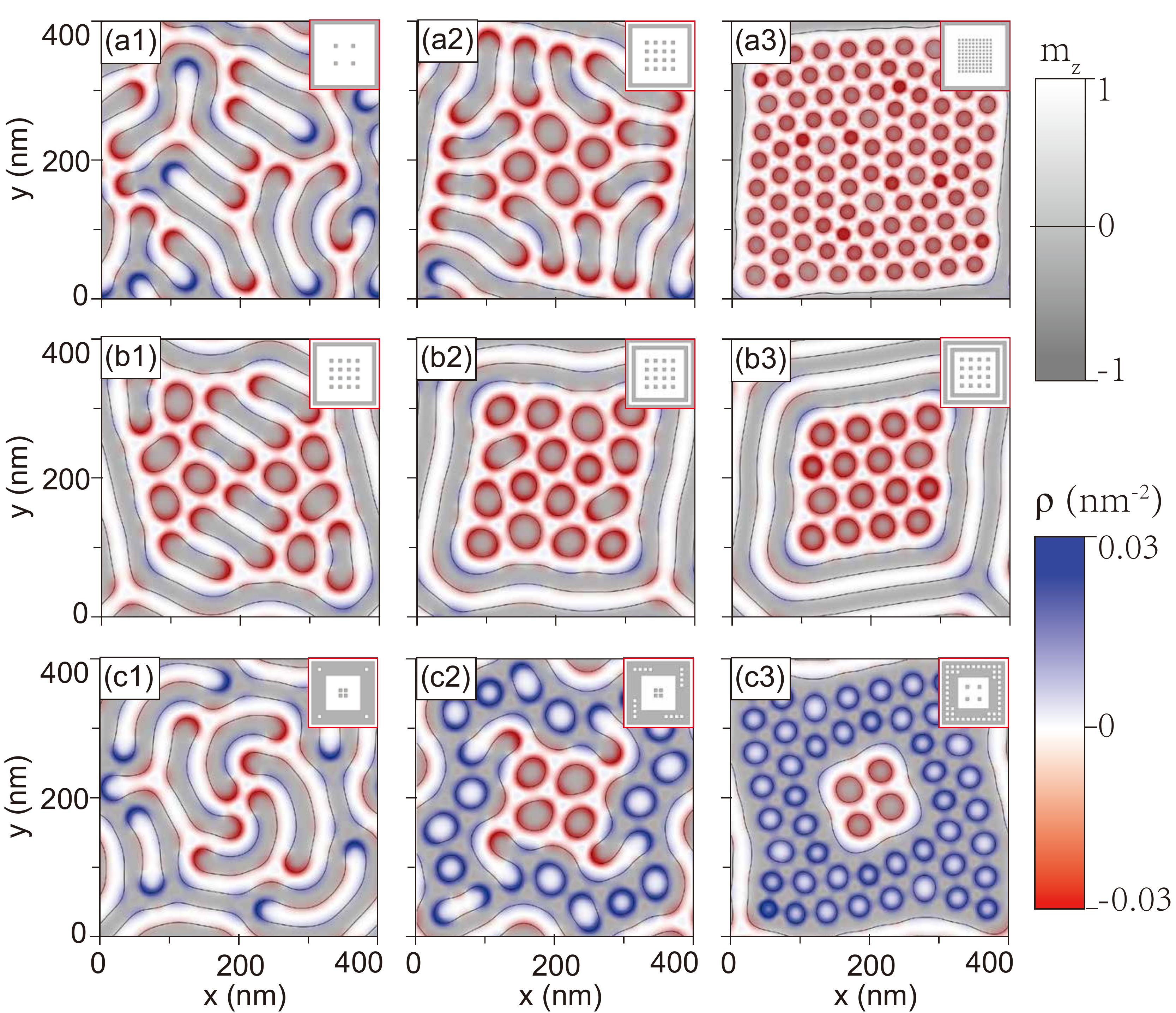}
\caption{Morphology and structures of various stable cascade stripe skyrmion bags 
for $\kappa=4$. (a1)-(a3) Stripe skyrmion bags of $n=2$ with $N=4$ (a1), 16 (a2) 
and 100 (a3) skyrmions in the innermost bag. (b1)-(b3) Cascade stripe skyrmion 
bags of $n=3$ (b1), 4 (b2) and 5 (b3) with $N=16$ skyrmions in the innermost bag. 
(c1)-(c3) Cascade stripe skyrmion bags of $n=3$ with $N=4$ skyrmions in the 
innermost bag and $4$ skyrmions (c1), 16 skyrmions (c2), and 36 skyrmions (c3) in 
the next innermost bag. Insets are the initial configurations of these cascade 
stripe skyrmion bags. Color bar denotes skyrmion charge density and the gray-scale 
encodes $m_z$.}
\label{fig7}
\end{figure*}

Skyrmions are stripes with negative formation energy when $\kappa>1$, and 
fundamental stripe width is $a(\kappa)2\pi A/D$, where $a(\kappa)$ quickly and 
monotonically decreases from $+\infty$ at $\kappa=1^+$ to 1 at $\kappa\gg 1$ 
\cite{paper1,paper2}. Thus, low energy states of a chiral magnetic film are 
condensed stripe skyrmions with the optimal stripe width. Different from the 
isolated skyrmions that cost energy to create skyrmion walls, stripes at low 
skyrmion density are flexible and irregular in order to fill up the film as 
shown in Figs. \ref{fig1}(b1)-(b3). Their widths should be resilient to 
cascade structures no matter whether they are in a single stripe skyrmion, 
or a target stripe skyrmion, or in a cascade stripe skyrmion bags, as long as 
the distance between two cascade layers are larger than the fundamental stripe 
width of $a(\kappa)2\pi A/D$. Figure \ref{fig5} shows stable target stripe 
skyrmions of $n=4$, 6 and 8 layers in a sample of $600\,\rm nm\times600\,
\rm nm \times 0.5\,\rm nm$ for $\kappa=1.1$ (a1)-(a3) and $\kappa=4$ (b1)-(b3). 
They are generated from the initial configurations shown in the insets of 
(a1)-(a3) for $n=4$, 6, and 8, respectively. The reasons for using these initial 
configurations are the same for those in Fig. \ref{fig1}. The average available 
layer space length of the target skyrmions are $600\,\rm nm/(2\times4)=75\,\rm nm$ 
for Figs. \ref{fig5}(a1) and \ref{fig5}(b1), $50\,\rm nm$ for Figs. \ref{fig5}(a2) 
and \ref{fig5}(b2), and $37\,\rm nm$ for Figs. \ref{fig5}(a3) and \ref{fig5}(b3), 
larger than the elementary stripe width of $L=32.5\, \rm nm$ for $\kappa=1.1$ 
and $L=25.10 \, \rm nm$ for $\kappa=4.$ Spin profiles of stripes in all 
layers are the same as that for elementary stripes as shown in Figs. 
\ref{fig5}(c)-(d). The symbols are numerical data along the green lines 
labelled by the green \textcircled{n} in Figs. \ref{fig5}(a1)-(a3) for (c) 
and along the orange lines labelled by the orange \textcircled{n} in Figs. 
\ref{fig5}(b1)-(b3) for (d). The solid curves are the fit to $\cos\Theta(x)=-
\frac{\sinh^2(L/2w)-\sinh^2(x/w)}{\sinh^2(L/2w)+\sinh^2(x/w)}$ for 
$-L/2<x<L/2$ and $\cos\Theta (x)=\frac{\sinh^2(L/2w)- \sinh^2\left[
(x-L/2-L/2)/w\right]}{\sinh^2(L/2w)+\sinh^2\left[(x-L/2-L/2)/w
\right]}$ for $L/2<x<L/2+L$ with $L=32.50\,$nm and $w=5.2\,$nm 
[Fig. \ref{fig5}(c)], and with $L=25.10\,$nm and $w=7.6\,$nm 
[Fig. \ref{fig5}(d)]. 

Different from the case of $\kappa<1$, the number of layers of a stable 
target stripe skyrmion can be any value in the thermodynamic limit, and has a 
maximal value for a finite sample. The maximal number is slightly larger than 
the half of the ratio between the sample size and the elementary stripe width. 
Figures \ref{fig6}(a1)-(a3) are three stable target stripe skyrmions for 
$\kappa=1.1$ with maximal layers of $n_{max}=4$, 8 and 11 when the sample 
sizes are $200\,\rm nm\times 200\,\rm nm \times0.5\,nm$, $400\,\rm nm\times400\,
\rm nm \times0.5\,nm$ and $600\,\rm nm\times 600\,\rm nm \times0.5\,nm$ for 
$\kappa=1.1$, respectively. They are from the initial configurations shown in the 
insets of corresponding figures with more layers than their final stable values. 
Similar to the evolutions of those stable circular target skyrmions in Fig. 
\ref{fig2}, the extra layers shrink and disappear one by one from the sample center. 
$n_{max}$ is very close to $200/(2\times 32.5)=3.15$; $400/(2\times 32.5)=6.3$;  
and $600/(2\times 32.5)=9.45$, corresponding to average available space size 
of $25\, \rm nm$ (a1), $25\, \rm nm$ (a2), and $\approx 27\,\rm nm$ (a3). 
The general features do not depend on the sample shape. As shown in Figs. 
\ref{fig6}(b1)-(b3) for rectangular samples of $200\,\rm nm\times400\,\rm nm
\times0.5\,nm$, $400\,\rm nm\times800\,\rm nm \times0.5\,nm$ and $600\,\rm nm
\times 1200\,\rm nm \times0.5\,nm$ with exactly the same model parameters 
of $\kappa=1.1$ as those in Figs. \ref{fig6}(a1)-(a3). $n_{max}=5$ (b1), 8 (b2), 
and 12 (b3) are mainly determined by the shorter length, and are similar to 
those in (a1)-(a3). Remarkably, the skyrmion in the innermost layer becomes 
a circular skyrmion of size $21 \,$nm (a1), $10.3\,$nm (a2) and $14.5\,$nm (a3), 
much smaller than the elementary stripe width of $L=32.5\,$nm. 
However, stripe width and spin profile in the outer layers are the same as 
the elementary stripes as shown in Fig. \ref{fig6}(e) with the symbols from the 
numerical simulations along the green lines labelled by the green \textcircled{n} 
in Fig. \ref{fig6}(b1)-(b3) and the solid curves from the fit to $\cos\Theta(x)=-
\frac{\sinh^2(L_1/2w)-\sinh^2(x/w)}{\sinh^2(L/2w)+\sinh^2(x/w)}$ for 
$-L/2<x<L/2$ and $\cos\Theta (x)=\frac{\sinh^2(L/2w)- \sinh^2\left[(x-L)/w\right]
}{\sinh^2(L/2w)+\sinh^2\left[(x-L)/w\right]}$ for $L/2<x<3L/2$ with $L=32.50\,$nm 
and $w=5.2\,$nm.

To show that this feature is general for all $\kappa>1$, Figs. \ref{fig6}(c1)-(c3) 
are three target stripe skyrmions with $n_{max}=5$, 9 and 13 in $200\,\rm nm\times 
200\,\rm nm \times 0.5\,nm$, $400\,\rm nm\times 400\,\rm nm \times 0.5\,nm$ and 
$600\,\rm nm\times 600\,\rm nm \times 0.5\,nm$ samples for $\kappa=4$, respectively. 
Since the elementary stripe width is reduced to $L=25.1\,$nm for $\kappa=4$, 
$n_{max}$ increases slightly and is again slightly larger than the ratio of sample 
size to $2L$ that is $\simeq 4$ (c1), 8 (c2) and 12 (c3). 
Figures \ref{fig6}(d1)-(d3) are the stable target stripe skyrmions with the maximal 
layers of $n_{max}=5$, 10 and 14 for rectangular samples of $200\,\rm nm
\times 400\,\rm nm\times0.5\,nm$, $400\,\rm nm\times800\,\rm nm \times0.5\,nm$ and 
$600\,\rm nm\times 1200\,\rm nm \times0.5\,nm$, respectively, with exactly the same 
model parameters as those in Figs. \ref{fig6}(c1)-(c3). Again, the skyrmion in the 
innermost layer is a circular object of $5.73 \,$nm (d1), $3.76 \,$nm (d2) and 
$6.44\,$nm (d3) in diameter that is smaller than the elementary stripe width 
of $L=25.1\,$nm. Stripe width and spin profile in the outer layers are the same as 
the elementary stripes as shown in Fig. \ref{fig6}(f) with the symbols from the 
numerical simulations along the orange lines labelled by the orange \textcircled{n} 
in Figs. \ref{fig6}(d1)-(d3) and the solid curves from the fit to $\cos\Theta(x)=-
\frac{\sinh^2(L_1/2w)-\sinh^2(x/w)}{\sinh^2(L/2w)+\sinh^2(x/w)}$ for $-L/2<x<L/2$ 
and $\cos\Theta (x)=\frac{\sinh^2(L/2w)- \sinh^2\left[(x-L)/w\right]}{\sinh^2(L/2w)
+\sinh^2\left[(x-L)/w\right]}$ for $L/2<x<3L/2$ with $L=25.1\,$nm and $w=7.63\,$nm. 

Figures \ref{fig5} and \ref{fig6} show clearly that stripes in all kinds of 
composite stripe skyrmions keep the properties of elementary stripe skyrmions,
such as their widths and spin profiles, at low skyrmion density such that the 
distance between two neighbouring layers is larger than the elementary width. 
Naturally, one interesting question is what will happen to the structure and 
morphology of composite skyrmions when more skyrmions are added into one 
particular layer or several layers. From the fact that a collection of elementary
stripe skyrmions form a SkX when the skyrmion-skyrmion distance is comparable 
to elementary stripe width \cite{paper3,paper4}, one expects that highly 
irregularly arranged flexible stripes of well-defined width as shown in Fig. 
\ref{fig7}(a1) deform their shapes to circular objects when the mean 
skyrmion-skyrmion distance inside one particular bag is reduced, and 
they eventually form SkXs in the layer(s) when the distance is around 
$L=a(\kappa)2\pi A/D$. Figure \ref{fig7} shows morphology change of cascade 
stripe skyrmions bags of various layers $n$ and $\kappa=4$ when the number 
of skyrmions in the innermost and next innermost bags increase from a low 
skyrmion density to a higher skyrmion density in a sample of $400\,\rm nm
\times 400\,\rm nm \times 0.5\,nm$. Figures \ref{fig7}(a1)-(a3) are the stable 
cascade stripe skyrmion bags of $n=2$ with $N=4$ (a1), 16 (a2), and 100 (a3) 
syrmions in the innermost bag. Skyrmions in the innermost bag are highly 
irregular stripes when the skyrmion density is low, as shown in Fig. \ref{fig7}(a1).  
It becomes a mixture of 4 circular skyrmions and 12 short straight stripe skyrmions 
when the average skyrmion-skyrmion distance is around $\sqrt{(400-2\times
25.10)^2/16}/2 \approx 44\,\rm nm \simeq 1.8 L$ as shown in Fig. \ref{fig7}(a2). 
100 skyrmions in the innermost bag for a nice triangular crystal as shown in Fig. 
\ref{fig7}(a3) when the average skyrmion-skyrmion distance is around $0.7 L$. 
Figures \ref{fig7}(b1)-(b3) are the stable cascade stripe skyrmion bags of $n=3$ 
(b1), 4 (b2) and 5 (b3) with $N=16$ skyrmions in the innermost bag. The bag size 
decreases as $n$ increases such that the effective skyrmion density in the bag 
increases. The 16 skyrmions form a mixture of circular and stripe skyrmions (b1), 
a SkX-like structure (b2), and an almost perfect SkX (b3). 
Figures \ref{fig7}(c1)-(c3) are cascade stripe skyrmion bags of $n=3$ with 
$N=4$ skyrmions in the innermost bag and 4 skyrmions (c1), 16 skyrmions (c2), 
and 36 skyrmions (c3) in the next innermost bag. When skyrmion densities in both 
bags are low, 4 skyrmions in both bags are irregular stripes as shown in (c1). 
Skyrmions in both bags are circular when the average skyrmion-skyrmion distance 
is around elementary stripe width as shown in (c2). Skyrmions in both bags form 
triangular lattice structures when the effective skyrmion densities are high 
enough as shown in (c3). 

\section{Discussion and conclusion}

It should be pointed out that many composite skyrmions discussed here 
were reported in previous micromagnetic simulations and in experiments. 
For example, skyrmionium were observed in space-coiling meta-structure 
\cite{skyrmionium6}, target skyrmions were observed in FeGe nanodisks 
\cite{skyrmionium4}, in space-coiling meta-structure \cite{skyrmionium6}, 
and in $\mathrm{Pd/Fe/Ir}$ hexagon systems\cite{skyrmionbag4}, more exotic 
cascaded skyrmion bags were observed in liquid crystals \cite{skyrmionbag1}, 
in $\mathrm{Fe_{3-x}Ge Te_2}$ \cite{skyrmionbag12}, and in chiral nematic 
liquid crystals \cite{ skyrmionbag13}. Here we propose a coherent way to 
organize these exotic skyrmion-like structures.

Composite skyrmions reveal fundamental nature of skyrmion duality. 
The duality allows all kinds of superstructures from elementary skyrmions, 
very similar to using bricks to construct different buildings in daily life. 
These superstructures provide countless new information carriers with 
distinct topologies and make new device concepts and new designs possible. 
For example, topology-based memory and gates should have high stability, high 
error tolerance and low noise
\cite{skyrmionbag6,skyrmionbag7,skyrmionbag8,skyrmionbag9,skyrmionbag10}. 
Topologically protected particle can also find 
its applications in electro-optic, microfluidic, and nanoparticle transport
\cite{skyrmionbag1,skyrmionbag11,skyrmionbag12,skyrmionbag13}. 
Charge and spin in solid state are coupled. Electron transport can be 
affected by the magnetic texture either through the emerged magnetic and 
electric fields, or simply through spin-orbit interactions. Thus, it should 
be highly interesting to identify those localized composite skyrmions which 
can be easily made on-demand, high controllability, and easily distinguish 
one from the others, especial electronically. 

Composite skyrmions can be used for magnetic data storage due to high stability 
and topological protection. They can also be used as magnetic sensors due to 
their sensitivity to external magnetic fields\cite{skyrmionbag11}. 
Another possible application of 
composite skyrmions is in energy-efficient computing applications, such as 
low-power spintronic devices and neuromorphic computing due to their varieties 
in morphology huge number of accessible states such that they can emulate the 
behavior of neurons\cite{Neurocomputer1}. One may use composite 
skyrmions to generate spin currents, 
which can be used to control magnetization. 

Although we did not consider magnetic field effect in this study, 
external magnetic fields can surely modify magnetic anisotropy that, 
in turn, can affect stability and properties of composite skyrmions\cite{skyrmionbag12}. 
Magnetic fields are surely an effective control knob. It should be 
very important and interesting to understand in detail how 
a magnetic field affects composite skyrmions, not only in terms of 
practical applications but also fundamentals, 

In conclusion, topology and static properties of stable/metastable magnetic 
textures of chiral magnets in the absence of magnetic field are fully 
determined by parameter $\kappa$, which measures relative strength of chiral 
interaction to the ferromagnetic exchange interaction and anisotropy, not 
by individual material parameters. $L=4A/(\pi D)$ is the fundamental 
length scale of all possible magnetic structures in a chiral magnet. 
Skyrmions have particle-continuum-medium duality that results in various 
interesting composite skyrmions such as target skyrmions and skyrmion bags. 
The complexities and the maximal number of cascade layers of a skyrmion 
superstructure with a given topology is sensitive to value of $\kappa$ 
for $\kappa\leq 1$ which support isolated circular skyrmions. 
For chiral magnetic films of $\kappa> 1$, where condensed stripe 
skyrmions is the thermodynamic equilibrium state at finite temperature, 
cascade stripe skyrmions maintain the width of elementary stripe skyrmions 
as long as the space between two cascade layers is larger than the width 
of elementary stripe skyrmions. When the space between two cascade layers is 
less than the fundamental stripe width, the cascade stripe skyrmions tend to 
be unstable. In the case of inserting many skyrmions inside one particular 
layer, whether in the innermost layer or any other cascade layer, such that 
the average skyrmion-skyrmion distance in around the fundamental stripe 
width, skyrmions because circular and form a triangular SkX. Our findings 
open the possibilities of new concepts and new designs in applications.   

\begin{acknowledgments}
This work is supported by the National Key Research and Development 
Program of China (No. 2020YFA0309600), the NSFC Grant (No. 11974296), and 
Hong Kong RGC Grants (No. 16301619, 16300522, and 16302321). 
\end{acknowledgments}

\appendix 

\section{Evidences of structure determination by $\kappa$ }

\begin{figure*}[htbp]
\centering
\includegraphics[width=17cm]{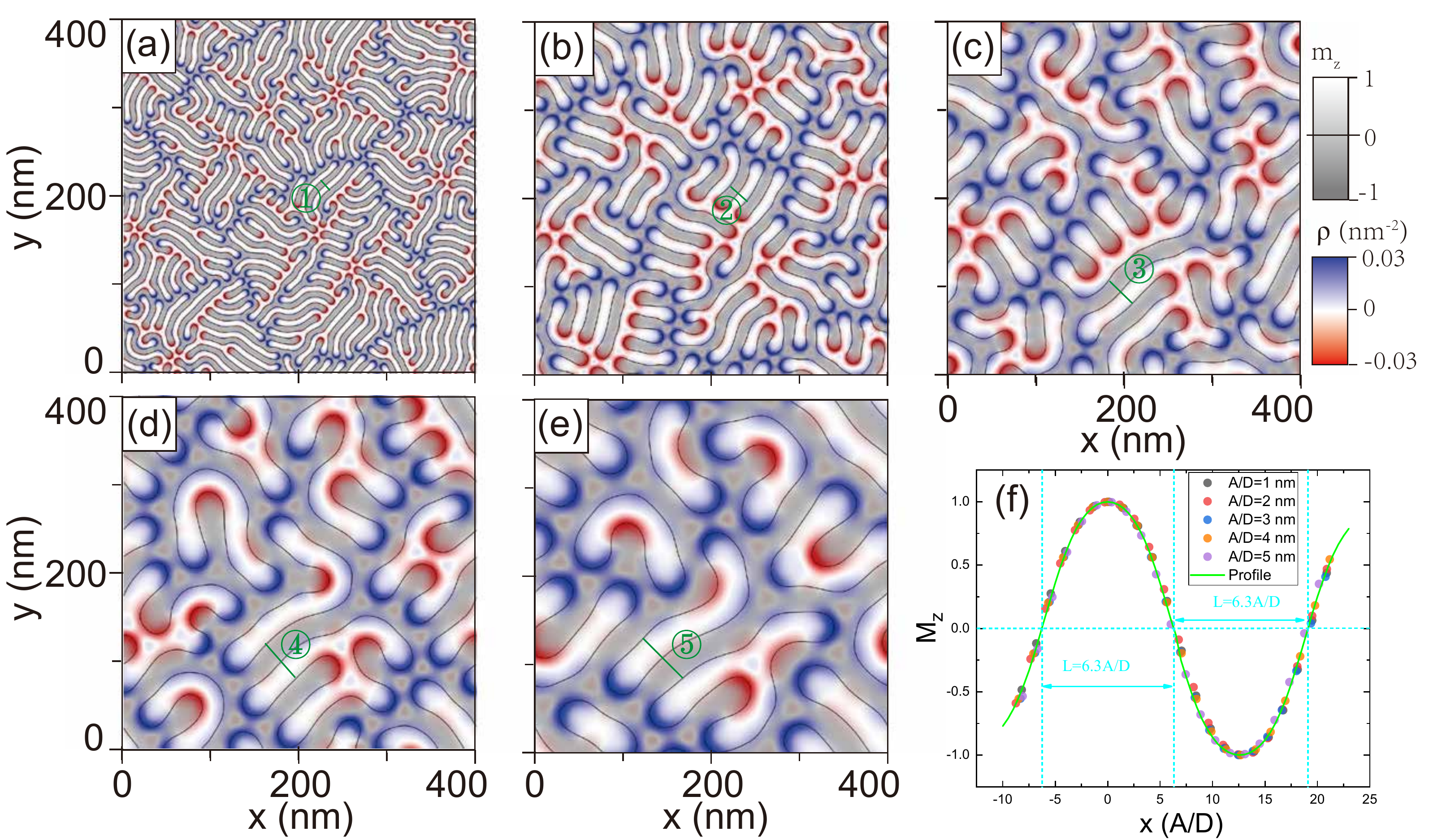}
\caption{(a)-(e) The structure of final state when A/D=1$\,$nm (a); 2$\,$nm (b); 
3$\,$nm (c); 4$\,$nm (d); 5$\,$nm (e). The inset of (a) shows the structure of the initial state. 
(f) The Spin profiles along the green lines labelled by \textcircled{1}-\textcircled{5} 
in (a)-(e). The symbols are numerical data and the solid lines are the theoretical fits 
to $M_z(x)=\frac{\sinh^2(L/w)-\sinh^2(x/2w)}{\sinh^2(L/w)+\sinh^2(x/2w)}$ for 
$-L/2\textless x \textless L/2$ and $M_z(x)=\frac{\sinh^2\left[(x-L)/2w\right]-\sinh^2(L/w)}
{\sinh^2(L/w)+\sinh^2\left[(x-L)/2w\right]}$ for $L/2\textless x \textless 3L/2$.
}
\label{figS}
\end{figure*}

The putative assertion from Eq. (\ref{sseq}) is that (meta)stable spin structures 
are completely determined by $\kappa$ up to a proper scaling is a sweet 
surprising. In order to further confirm this assertion, we use a film 
of $400\,$nm$\times 400\,$nm$\times 0.5\,$nm described by Eq. (\ref{energy}) to 
simulate various metastable structures with the same $\kappa$ and different 
$A/D$. We set $\kappa=4$ that supports condensed stripe skyrmions and with 
10 nucleation domain randomly distributed in the film. Initially, the model 
parameters are $A=1\, \rm pJ\, m^{-1}$, $D=1\, \rm mJ\, m^{-2}$, 
$K_u=0.154\, \rm MJ\, m^{-3}$, $M_s=0.58\,  \rm MA\, m^{-1}$. 
Figure \ref{figS}(a) below shows the morphology of the final stable 
structure with 10 ramified stripe skyrmions. Then we gradually increase 
$A/D$ from $1\,$nm to $5\,$nm and decrease $K_u$ to keep the $\kappa=4$
unchanged, and Figs. \ref{figS}(b)-(e) are the final stable 
structures for $A/D= 2\,$nm (b); $3\,$nm (c); $4\,$nm (d); $5\,$nm (e). 
Although these structures are visually different, showing multiply 
solutions of Eq. (\ref{sseq}), they are in fact similar. Fig. \ref{figS}(f) plot $M_z (x)$ 
along the green lines labelled by \textcircled{1}-\textcircled{5} in Figs. 
\ref{figS}(a)-(e). The spin profiles fall on the same curve when $x$ is plotted in 
the units of $A/D$. The results are exactly what the assertion is expected. 

\end{document}